\documentclass[twocolumns,useAMS,articles,fleqn] {mnras}
\usepackage{epsfig}
\usepackage{amsmath}
\usepackage{graphicx}
\usepackage{amssymb}

\title[Structure of massive star forming regions]{The evolution of the internal structure of massive star forming regions in the Milky Way as revealed by ALMA}

\author[Dib]{Sami Dib$^{1}$\thanks{E-mail, SD: sami.dib@gmail.com; dib@mpia.de}\\
$^{1}$Max Planck Institute for Astronomy, K\"{o}nigstuhl 17, 69117, Heidelberg, Germany\\
}

\begin{document}
\maketitle

\date{Accepted XXX. Received XXX}

\pagerange{\pageref{firstpage}--\pageref{lastpage}}
\pubyear{2016}
\label{firstpage}

\begin{abstract} 
We analyze the structure of 15 protocluster forming regions in the Milky Way using their $1.3$ mm continuum emission maps from the ALMA-IMF large program. The analysis of the clouds structure is performed using the delta-variance spectrum technique. The calculated spectra display a self-similar regime on small scales as well as the presence of a prominent bump on larger scales and whose physical size, $L_{\rm hub}$, falls in the range of $\approx 7000$ au to $60000$ au. These scales correspond to the sizes of the most compact clumps within the protocluster forming clouds. A significant correlation is found between $L_{\rm hub}$ and the surface density of the free-free emission estimated from the integrated flux of the H41$\alpha$ recombination line $\left(\Sigma_{\rm H41\alpha}^{\rm free-free}\right)$ as well as a significant anti-correlation between $L_{\rm hub}$ and the ratio of the 1.3 mm to 3 mm continuum emission fluxes $\left(S_{\rm 1.3 mm}^{\rm cloud}/S_{\rm 3 mm}^{\rm cloud}\right)$. Smaller values of $\left(S_{\rm 1.3 mm}^{\rm cloud}/S_{\rm 3 mm}^{\rm cloud}\right)$ and larger values of $\Sigma_{\rm H41\alpha}^{\rm free-free}$ correspond to more advanced evolutionary stages of the protocluster forming clumps. Hence, our results suggest that the sizes of the densest regions in the clouds are directly linked to their evolutionary stage and to their star formation activity with more evolved clouds having larger protocluster forming clumps. This is an indication that gravity plays a vital role in regulating the size and mass growth and star formation activity of these clumps with ongoing gas accretion.
\end{abstract} 


\section{INTRODUCTION}\label{introduction}

Massive stars play a vital role in the galactic ecosystem, both in the Milky Way and in external galaxies. Feedback from massive stars contributes a significant fraction of the galaxy's energy budget (Mac Low \& Klessen 2004) and regulate the dynamics of interstellar gas on kpc scales down to smaller scales ($\approx10$ pc) that are typical of the sizes of molecular clouds (Dib et al. 2006,2009; Bacchini et al. 2020; Lu et al. 2020; Ganguly et al. 2022). Massive stars are responsible for expelling the gas from star forming regions and for setting the star formation efficiency (Dib et al. 2011,2013; Hony et al. 2015; Kim et al. 2021). They are also the primary channel for the galactic chemical enrichment as they return their processed heavy elements into the galactic interstellar medium during their lifetimes in the form of stellar winds or at the time of their demise as supernova explosions (Izotov \& Thuan 2000; Nomoto et al. 2013). 

\begin{figure*}
\begin{center}
\includegraphics[width=0.36\textwidth] {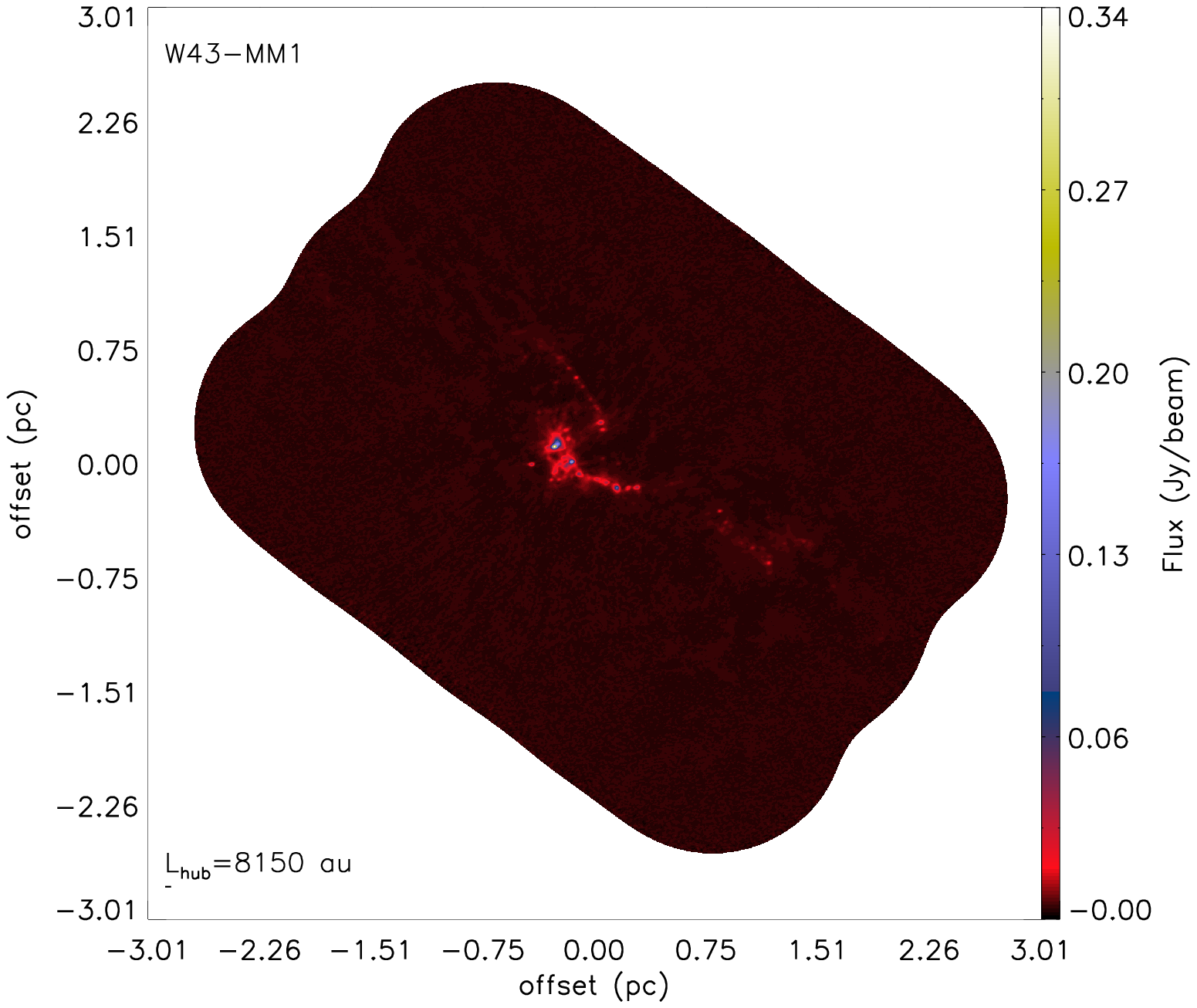}
\hspace{3cm}
\includegraphics[width=0.36\textwidth] {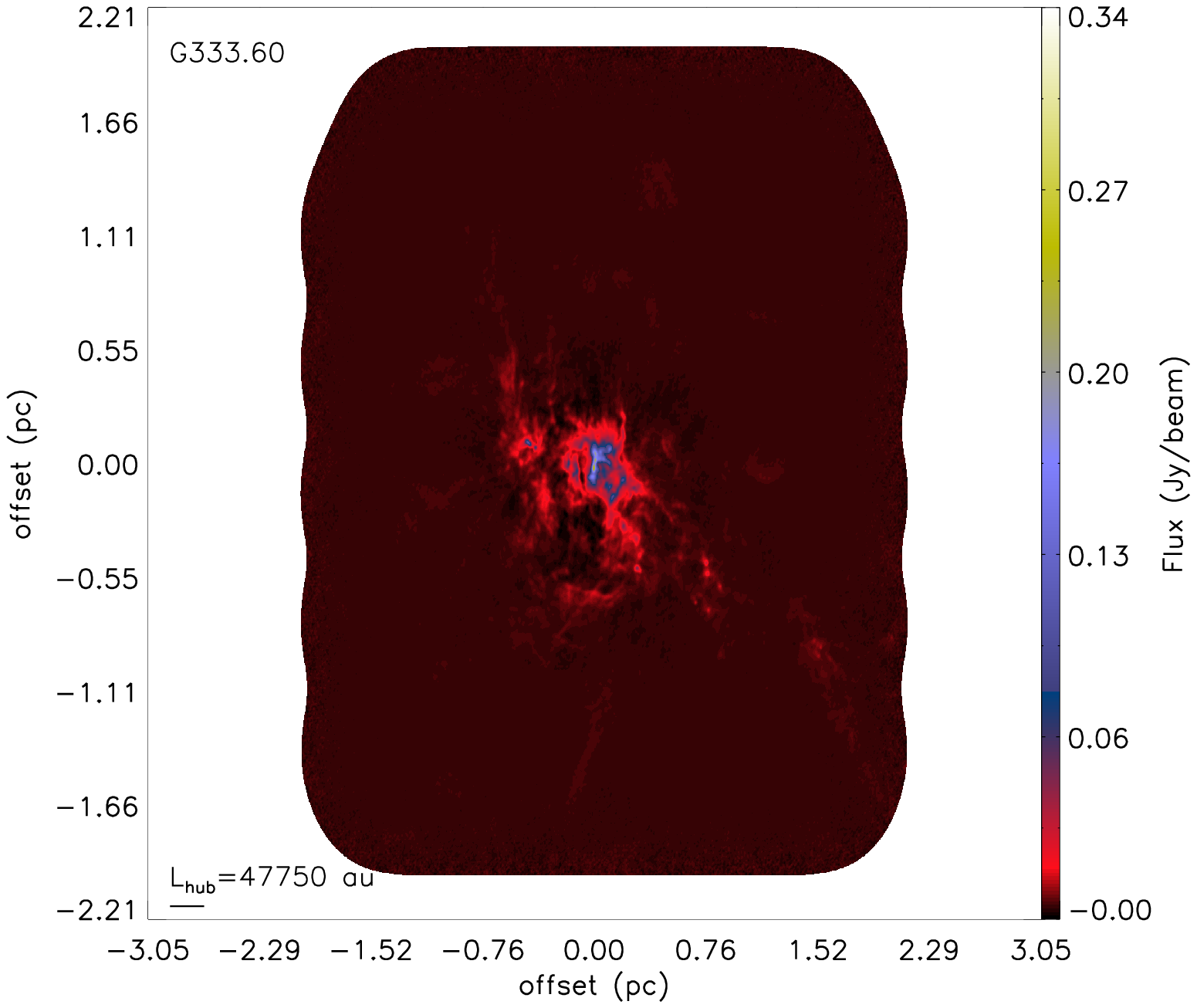}
\vspace{0.75cm}
\caption{The 1.3 mm continuum maps of two selected regions from the total of 15 regions mapped by the ALMA-IMF project. They correspond to a region classified as Young (left panel), and Evolved (right panel). The labels on each axis display the offset from the center of the map. In the lower left corner of each figure, the spatial scale corresponding to the value of $L_{\rm hub}$ is displayed. The remaining 13 other regions are not shown and can be viewed in the original papers (Motte et al. 2022; Ginsburg et al. 2022).}
\label{fig1}
\end{center}
\end{figure*}  

The formation of massive stars is a complex process that begins with the formation of their parental molecular clouds, the fragmentation of the clouds, and the formation of massive protostars (Zinnecker \& Yorke 2007; Dib \& Henning 2019). The cloud assembly process can find its origin in generic thermal and gravitational instabilities (Dib \& Burkert 2004,2005) or by the collision of lower mass clouds (Dobashi et al. 2019; Fujita et al. 2021; Uchiyama et al. 2022). The fragmentation of massive clouds is regulated by competition between gravity and magnetic fields and the turbulent motions of the gas inherited from the cloud assembly phase as well as those generated by protostellar jets from the newly formed protostars (Cortes et al. 2008; Csengeri et al. 2011; Beuther et al. 2015,2019; Cunningham et al. 2018; Li et al. 2019; Liu et al. 2020; Liu et al. 2022; Beltr\'{a}n et al. 2022). As massive stars evolve quickly towards the main sequence, their strong radiation fields and stellar winds further impact the cloud structure and dynamical state (Gritscnheder et al. 2009; Tremblin et al. 2012; Dale et al. 2014; Carlsten \& Hartigan 2018; Nayak et al. 2021; Geen et al. 2021; Rebolledo et al. 2020; Tiwari et al. 2021). 

\begin{figure*}
\begin{center}
\includegraphics[width=0.9\textwidth] {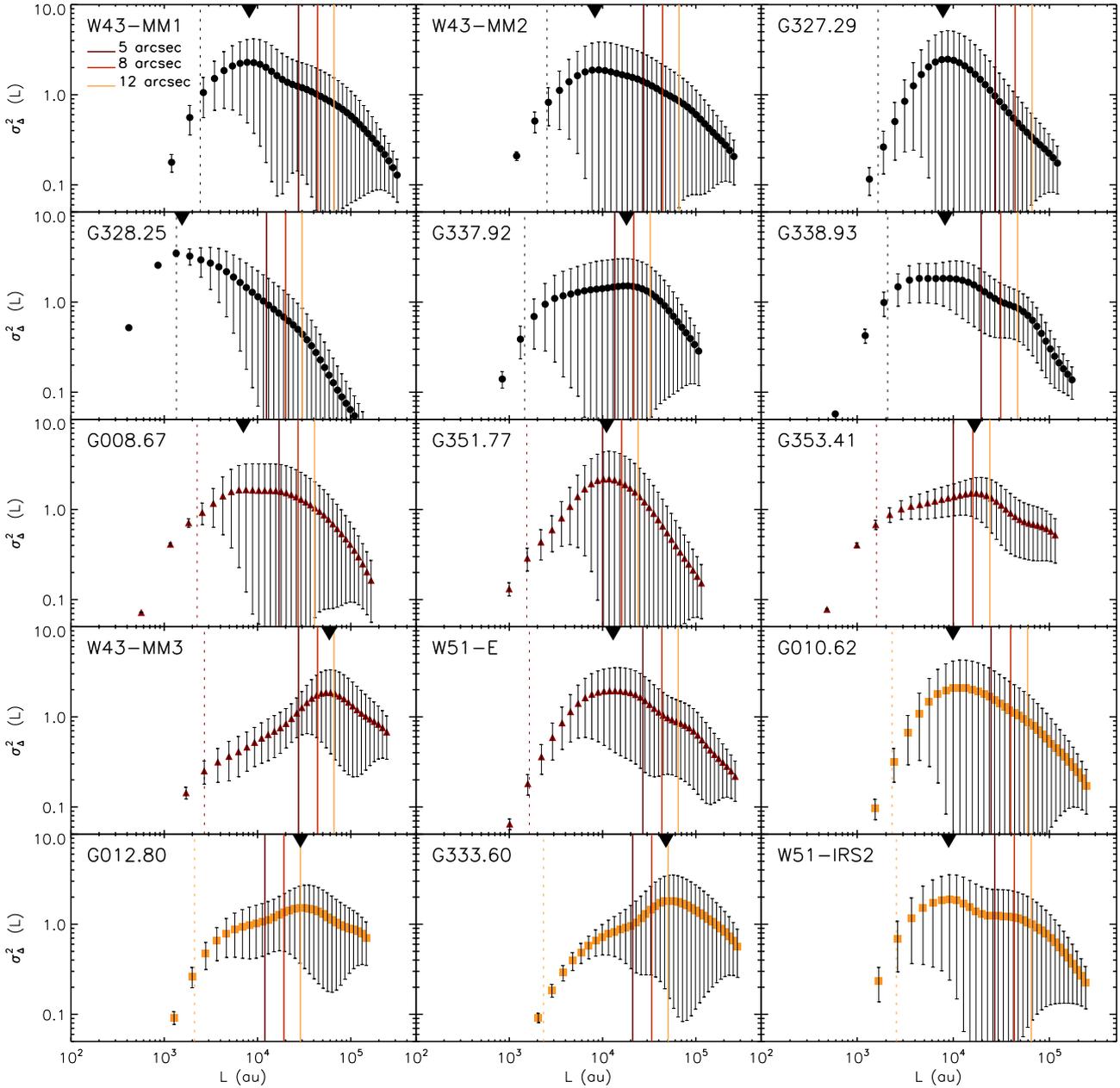}
\vspace{0.75cm}
\caption{The $\Delta$-variance spectra of the 15 star forming regions in the ALMA-IMF sample. The spectra in black correspond to regions classified as Young, those in dark red to those classified as Intermediate, and ones in yellow to those classified as Evolved (see text for details). The vertical dashed lines corresponds to the spatial resolution of the beam. All spectra are normalized by their respective mean value. The vertical arrow in each panel marks the position of the peak in the spectrum. Also shown in each spectrum are the spatial scales corresponding to angular resolutions of $5\arcsec$ (full brown line), $8\arcsec$ (full red line), and $12\arcsec$ (full yellow line). These values bracket the value of the MRS which is suspected to fall in this range.}
\label{fig2}
\end{center}
\end{figure*}

In terms of global structure, massive star forming regions are often observed to have a hub-filament system architecture (e.g., Elia et al. 2018; Dib et al. 2020; Dewangan 2022; Zhou et al. 2022) with most, if not all protostars residing in the hub substructure (Kumar et al. 2020). In analyzing the global morphology of massive star forming regions and particularly of the hub+filament systems, it is customary, and often very useful, to decompose the entire structure into a hub and a series of individual filaments in order to study both their individual and statistical properties (Kumar et al. 2020). A complementary approach in which the structure of the star-forming region as a whole is quantified can help shed light on how the total emission is distributed across different physical scales. In this work, we perform such an analysis on 15 massive star-forming regions in the Milky Way using the $1.3$ mm continuum emission maps from the ALMA-IMF project. We quantify the internal structure of these star-forming regions using the delta-variance ($\Delta$-variance) spectrum technique and we explore if and how features in the spectra relate to the clouds' evolutionary stages. The data is presented in Sect.~\ref{obsdata} and the calculation of the $\Delta$-variance is briefly summarized in Sect.~\ref{deltavar}. The results and conclusions are presented in Sect.~\ref{res} and Sect.~\ref{conc}, respectively.

\section{DATA}\label{obsdata}

We used data from the ALMA-IMF\footnote{\url{https://www.almaimf.com/data.html}} project which is a large program of the {\it Atacama Large Millimeter Array} (ALMA) (Motte et al. 2022; Ginsburg et al. 2022). The sample consists of $15$ massive star forming regions located a distances between $\approx 2$ kpc to $5.5$ kpc and that are in different evolutionary stages. Out of the 15 protocluster regions, six are categorized as being Young (Y), five as Intermediate (I), and four as Evolved (E). This classification in terms of evolutionary stage is based on two criteria which are the 1.3 mm to 3 mm flux ratio and the free-free emission at the H41$\alpha$ frequency of the recombination line. The adoption of these criteria is based on the assumption that as massive star forming regions evolve, they will host, statistically, an increasing number of \ion{H}{ii} regions. As these \ion{H}{ii} regions expand, their free-free emission increases. Furthermore, for evolved clouds, the free-free emission dominates at 3 mm, whereas emission at the 1.3 mm emission is dominated by thermal dust emission. This leads to ratios of the $1.3$ mm to $3$ mm emission that decrease as clouds evolve. Motte et al. measured the ratio of the $1.3$ mm and $3$ mm using the surface area of the $1.3$ mm emission for each cloud $\left(S_{\rm 1.3 mm}^{\rm cloud}/S_{\rm 3 mm}^{\rm cloud}\right)$ and they also made an estimate of the surface density of the free-free emission $\left(\Sigma_{\rm H41\alpha}^{\rm free-free}\right)$. Table~\ref{tab1} lists the values of $\left(S_{\rm 1.3 mm}^{\rm cloud}/S_{\rm 3 mm}^{\rm cloud}\right)$ and $\Sigma_{\rm H41\alpha}^{\rm free-free}$ measured by Motte et al. (2022).

For each protocluster, the current data products available from ALMA-IMF are two versions of a $1.3$ mm and a $3$ mm continuum maps which were obtained using the 12m array configuration of ALMA (Ginsburg et al. 2022). The first version (bsens) is more sensitive, but is affected by emission lines present in each selected bandwidth. The second type of maps (cleanest) are built using the line-free channels. In this work, we will focus the analysis on the cleanest $1.3$ mm maps (the image.tt0.pbcor versions). For these maps, the spatial resolution varies from region to region and is in the range of $1350$ to $2690$ au (see Table 5 in Motte et al. 2022). In terms of the number of pixels, the size in pixels of the ALMA-IMF $1.3$ mm maps varies from $882\times882$ pixels up to $2048\times2048$ pixels, with the majority of maps having more than 1000 pixels in each direction. The $1.3$ mm maps of two regions from the entire ALMA-IMF sample are displayed in Fig.~\ref{fig1}, namely those of the Young region W43-MM1 and the Evolved region G333.60).    

\begin{figure}
\begin{center}
\includegraphics[width=0.9\columnwidth] {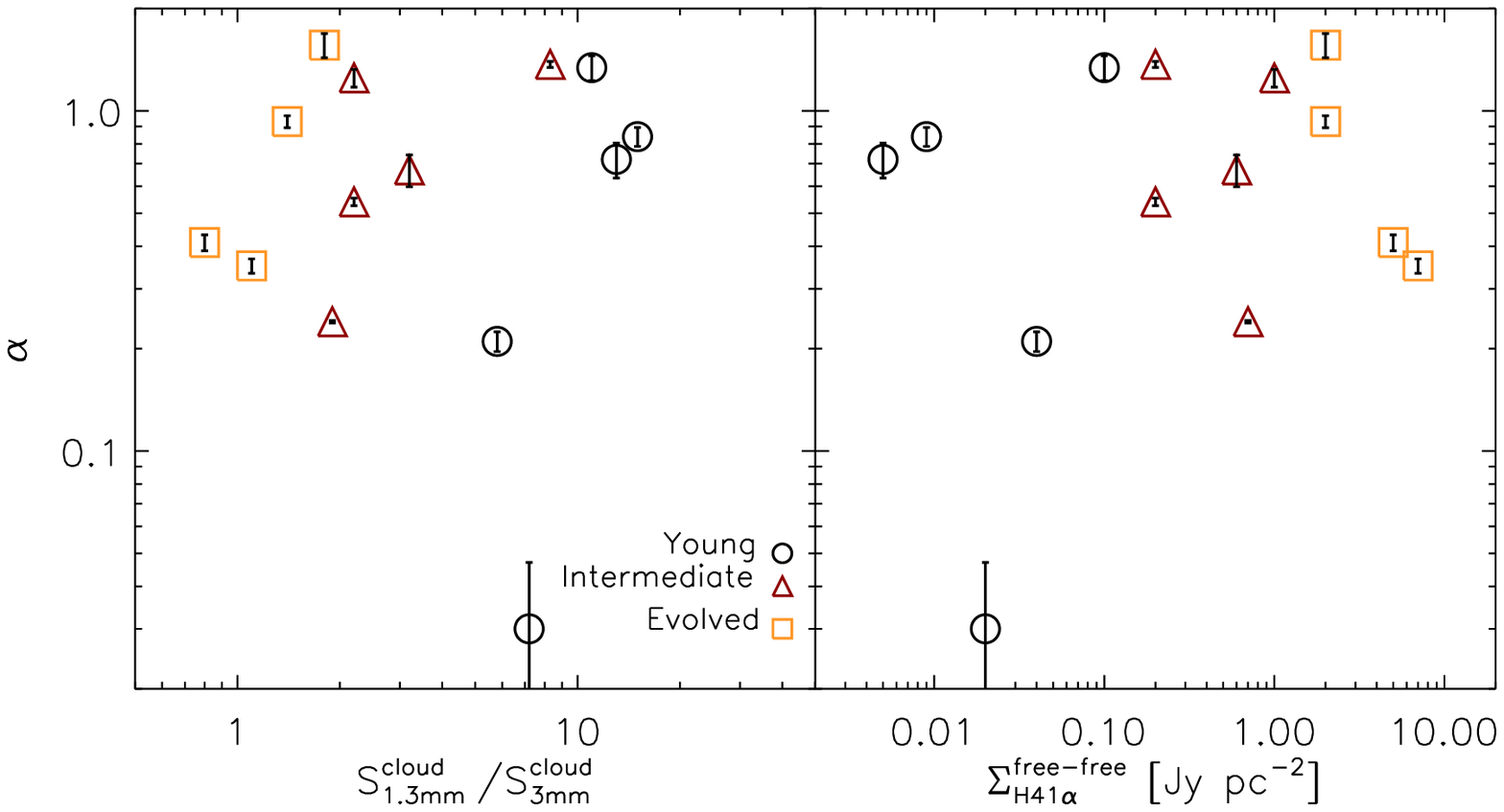}
\vspace{0.75cm}
\caption{Scatter plot between the slope of the $\Delta$-variance spectrum in the self-similar regime, $\alpha$, and the ratio $\left(S_{\rm 1.3 mm}^{\rm cloud}/S_{\rm 3 mm}^{\rm cloud}\right)$ (left panel) and between $\alpha$ and the surface density of the H41$\alpha$ free-free emission (right panel). The black circles are regions classified as Young, the dark red triangles correspond to those classified as Intermediate, and the yellow squares to those classified as evolved.}
\label{fig3}
\end{center}
\end{figure}  

\begin{figure}
\begin{center}
\includegraphics[width=0.9\columnwidth] {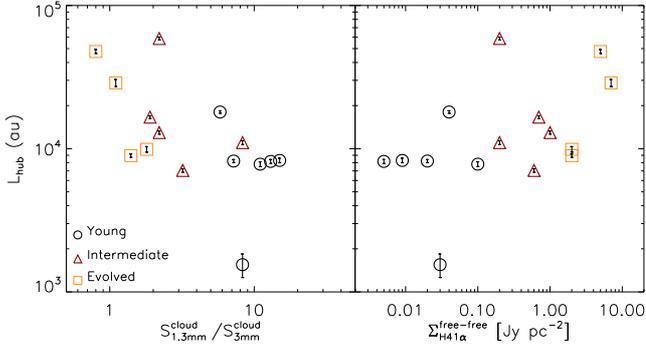}
\vspace{0.75cm}
\caption{Scatter plot between the peak found in the $\Delta$-variance spectrum, $L_{\rm hub}$, and the ratio $\left(S_{\rm 1.3 mm}^{\rm cloud}/S_{\rm 3 mm}^{\rm cloud}\right)$ (left panel) and between $\alpha$ and the surface density of the H41$\alpha$ free-free emission (right panel). The black circles are regions classified as Young, the dark red triangles correspond to those classified as Intermediate, and the yellow squares to those classified as evolved.}
\label{fig4}
\end{center}
\end{figure}

\section{ANALYSIS: THE $\Delta$-VARIANCE SPECTRUM}\label{deltavar}
We quantified the structure of the $1.3$ mm maps using the $\Delta$-variance spectrum technique. The method was introduced in Stutzki et al. (1998) and Zielinsky et al. (1999) and is a generalization of the Allan variance (Allan 1966). In this work, we used an improved implementation of the method presented in Ossenkopf et al. (2008a)\footnote{\url{https://hera.ph1.uni-koeln.de/~ossk/Myself/deltavariance.html}}. For a two-dimensional field  $A(x,y)$, the $\Delta$-variance on a scale $L$ is defined as being the variance of the convolution of $A$ with a filter function $\sun_{L}$ , such that

\begin{equation}
\sigma_{\Delta}^{2}(L)=\frac{1}{2\pi} \langle (A * \sun_{L})^{2}  \rangle_{x,y},
\label{eq1}
\end{equation}

\noindent where $x$ and $y$ represent the distances on both axis separating the point under consideration from any other point. For the filter function, Ossenkopf et al. (2008a) use a Mexican hat function which we also adopt here. This is defined as

\begin{equation}
\sun_{L} \left(r\right)= \frac{4}{\pi L^{2}} e^{\frac{r^{2}} {(L/2)^{2}}} - \frac{4}{\pi L^{2} (v^{2} -1)} \left[ e^{\frac{r^{2}}{(vL/2)^{2}}} -e^{\frac{r^{2}}{(L/2)^{2}}}\right],
\label{eq2}
\end{equation}

\noindent where the two terms on the right side of Eq.~\ref{eq2} represent the core and the annulus of the Mexican-hat function, respectively, and $v$ is the ratio of their diameters (we used a value of $v=1.5$) and where $r=\sqrt{x^{2}+y^{2}}$. For a faster and more efficient computation of Eq.~\ref{eq1} the calculation is performed as a multiplication in Fourier space (Ossenkopf et al. 2008a), and thus, the $\Delta$-variance is given by

\begin{equation}
\sigma_{\Delta}^{2}(L)=\frac{1}{2 \pi} \int \int P \left| \bar{\sun}_{L} \right |^{2} dk_{x} dk_{y},
\label{eq3}
\end{equation}  

where $P$ is the power spectrum of $A$, and $\bar{\sun}_{L}$ is the Fourier transform of the filter function. If $\beta$ is the exponent of the power spectrum ($P(k) \propto k^{-\beta}$, where $k=2\pi/L$ is the wave number), then a relation exists between the slope of the $\Delta$-variance and $\beta$ (St\"{u}tzki et al. 1998). This is given by

\begin{equation}
\sigma_{\Delta}^{2}(L) \propto L^{\alpha} \propto L^{\beta-2}.
\label{eq4}
\end{equation}

The slope of the $\Delta$-variance can be measured from the range of spatial scales over which it displays a self-similar behavior. It can be related to the value of $\beta$. Characteristic scales are scales at which there are breaks of the self-similarity and that appear in the $\Delta$-variance plots as break points or inflexion points. The error bars of the $\Delta$-variance are computed from the counting error determined by the finite number of statistically independent measurements in a filtered map and the variance of the variances, that is, the fourth moment of the filtered map. The $\Delta$-variance has been used to analyze the structure of observed molecular clouds (Bensch et al. 2001; Sun et al. 2006; Ossenkopf et al. 2008b; Russeil et al. 2013; Elia et al. 2014; Dib et al. 2020; Schneider et al. 2022) as well the structure of the \ion{H}{i} gas in nearby galaxies (Elmegreen et al 2001; Dib et al. 2021). In molecular clouds, a variety of features can be observed in the $\Delta$-variance spectrum (Russeil et al. 2013; Elia et al. 2014; Dib et al. 2020; Schneider et al. 2022). Low surface-density clouds which are not forming stars display a self-similar regime which can be described by a single power law while at the other extreme, massive clouds that are intensely forming stars display bumps in their spectra which are associated with the existence of characteristic scales of the order of $\approx 0.8-1$ pc (Dib et al. 2020). Dib et al. (2020) interpreted this scale as representing the typical size of hubs in hub+filaments systems observed at the resolution of the {\it Herschel} satellite.

\section{RESULTS}\label{res}

\begin{table*}
\begin{center}
\begin{tabular}{lccccc}
\hline
\hline
Protocluster cloud name &   $S_{\rm 1.3 mm}^{\rm cloud}/S_{\rm 3 mm}^{\rm cloud}$ &  $\Sigma_{\rm H41\alpha}^{\rm free-free}$ [Jy pc$^{-2}$]  &  $\alpha$  & $\alpha$-range [$10^{3}$ au] & $L_{\rm hub}$ [au]        \\ 
 \\
\hline
        W43-MM1 & 13  &  0.005 & 0.72   & [3-6] & 8150 \\
        W43-MM2 & 15  &  0.009 & 0.84  & [3-6] &  8280 \\
        G338.93    & 7.2 &  0.02   &  0.03   & [3-10]  & 8200 \\
        G328.25    & 8.3 &  0.03   & -   & -    & 1550 \\
        G337.92    & 5.8 & 0.04    &  0.21  & [3-10]   & 18040 \\ 
        G327.29    & 11  &  0.1     & 1.34   & [3-6]  & 7800 \\
        \hline
        G351.77    & 8.3 & 0.2 & 1.37 &  [3-6] &  11025 \\
        G008.67    & 3.2 & 0.6 & 0.67 &  [3-6] & 7050 \\
        W43-MM3 & 2.2 & 0.2  & 0.54  & [3-10]  & 58800  \\
        W51-E      & 2.2 & 1      & 1.25  & [3-6]&  12960 \\ 
        G353.41    & 1.9 & 0.7  &  0.24  & [3-15]  & 16600 \\ 
        \hline
        G010.62    & 1.8 & 2   & 1.56  & [3-6] & 9900 \\
        W51-IRS2  &1.4 &  2  &  0.93 & [3-6] & 8940 \\
        G012.80     & 1.1 & 7  &  0.35  & [6-20]  & 28800 \\ 
        G333.60    & 0.8 & 5   &  0.41    & [10-25]  & 47750\\
         \hline 
\end{tabular}
\caption{Characteristics of the 15 star forming regions in the ALMA-IMF sample and of their $\Delta$-variance spectra. Columns represent the (1) the name of the protocluster cloud, (2) the $1.3$ mm to $3$ mm flux ratio, measured on the area of the $1.3$ mm emission, (3) the surface density of the free-free emission, (4) the slope of the $\Delta$-variance spectra in the self-similar regime, (5) the spatial range over which $\alpha$ is measured, (6) the position of the peak of the $\Delta$-variance spectrum, $L_{\rm hub}$.}
\label{tab1}
\end{center}
\end{table*}

Figure~\ref{fig2} displays the $\Delta$-variance spectra of the 15 regions of the ALMA-IMF sample. All spectra are normalized by their mean value. The spectra are split into three sub-groups which represent the protocluster clouds classified as Young, Intermediate, and Evolved following the classification by Motte et al. (2022). The aim of this segregation is to verify whether the clouds' evolutionary stage has any connection to their internal structure. The first noticeable effect is that the spectra of all regions deviate from a single power law that is observed in low star forming regions (Dib et al. 2020; Yahia et al. 2021; Schneider et al. 2022). Instead, the $\Delta$-variance spectra of the ALMA-IMF regions resemble those calculated for massive star forming regions found at smaller distances and that have been obtained using data from the {\it Herschel} satellite such as NGC 6334 (Russeil et al. 2013), M17 (Schneider et al. 2022), and Cygnus X (Dib et al. 2020). The spectra are characterized by the presence of a bump on scales that range from several $10^{3}$ au to several $10^{4}$ au. Dib et al. 2020 presented an extensive suite of tests that shows how the presence of over-dense regions on top of a self-similar regime in two-dimensional images leads to the formation of a bump in the $\Delta$-variance spectrum (see figures 6-9 in their paper), a result that corroborates similar findings presented in Elia et al. (2014) and in particular in figure 3 of their paper. Thus, the bumps observed in the $\Delta$-variance spectra of the ALMA-IMF fields are associated with the presence of over-dense structures that constitute the central hub(s) of the star forming clouds and such features are not observed in less massive regions, especially those that do not show signs of ongoing star formation such as the Polaris flare region (Dib et al. 2020). Figure~\ref{fig2} also shows that in regions that are classified as Intermediate and Evolved, the peaks of the spectra are present at larger physical scales that those found in the less evolved, young regions (see Fig.~\ref{fig1} for examples, and also Tab.~\ref{tab1} for the $L_{\rm hub}$ values of the entire sample). Larger values of $L_{\rm hub}$ in Evolved regions are a clear indication that massive star forming regions such as those mapped in the ALMA-IMF project continue to accrete gas from their surroundings leading them to become larger and more massive\footnote{The masses of the clouds and of their main clumps (i.e., the hubs) have yet to be been estimated in the ALMA-IMF maps. However, their counterparts in ATLASGAL follow a mass-size relationship (Csengeri et al. 2017), and one can reasonably expect that a similar law exists for the $1.3$ mm maps as seen in the ALMA observations.}, and in turn, this induces stars to form at a higher rate. 

We now explore the connection between features in the $\Delta$-variance spectrum and the evolutionary criteria $\left(S_{\rm 1.3 mm}^{\rm cloud}/S_{\rm 3 mm}^{\rm cloud}\right)$ and $\Sigma_{\rm H41\alpha}^{\rm free-free}$. For each region, we extracted two sets of information from the $\Delta$-variance spectra. The first one is the slope of the spectra in the self-similar regime that precedes the bump on the left-hand side of the spectra. As shown in Dib et al. (2020), the self-similar regime in the clouds could be perturbed by the presence of a bump and at the smallest scales, it can be affected by the effects of beam smearing which can extend beyond the actual beam size by a factor of $\approx 2$. We adopted a conservative approach and fit the spatial ranges of the spectra that display a clear self-similar regime, avoiding perturbed scales at both ends. The ranges over which a power law are fit to the self-similar regimes are listed in Tab.~\ref{tab1} along with the derived values of the exponent $\alpha$. Figure~\ref{fig3} displays the values of $\alpha$ plotted against the ratio $\left(S_{\rm 1.3 mm}^{\rm cloud}/S_{\rm 3mm}^{\rm cloud}\right)$ (left panel) and $\Sigma_{\rm H\alpha{41}}^{\rm free-free}$ (right panel). The G328.25 cloud is not included as its spectrum does not show the presence of a self-similar regime on spatial scales below the peak. There is no correlation between $\alpha$ and $\left(S_{\rm 1.3 mm}^{\rm cloud}/S_{\rm 3mm}^{\rm cloud}\right)$ (Spearman $\rho$ coefficient of 0.16) and between $\alpha$ and $\Sigma_{\rm H\alpha{41}}^{\rm free-free}$ (Spearman $\rho$ coefficient of 0.10). The significance of the deviation from zero of these two coefficients is 0.56 and 0.73, respectively, which confirms the absence of any correlation between $\alpha$ and the star formation indicators. We have tested the robustness of this result by varying the ranges over which the power law fit for the self-similar regimes is performed. This was done by shifting the lower and upper bound of the range by 1000 au to larger scales, individually or simultaneously. Within the 1$\sigma$ uncertainties, the resulting exponents of the different power laws remain identical.

In principle, the position of the peak, $L_{\rm hub}$, can be estimated by eye. However, we derived a more accurate estimate of its value with a simple quadratic interpolation using several points around the peak. The values of $L_{\rm hub}$ are reported in Tab.~\ref{tab1} and are marked in Fig.~\ref{fig2} with a downward pointing arrow. There is no correlation between the measured values of $L_{\rm hub}$ and the distances of the regions that are reported in Motte et al. (2022). It is relevant to point out that the ALMA-IMF images used in this work are based on observations with the 12m array, and as such, some of the emission at the largest scales can be filtered out. The determined value of $L_{\rm hub}$ can be compared to the value of the maximum recoverable scale (MRS) for each field of the ALMA-IMF sample in order to assess whether the measurement of $L_{\rm hub}$ corresponds to real dense structures found in the clouds. We refer the reader to the original paper by Ginsburg et al. (2022) for a detailed discussion on what could affect the value of the MRS for those regions. The assessment of Ginsburg et al. (2022) is that the MRS varies from field-to-field and is difficult to quantify. Judging by the visibility coverage of the interferometer in the 12m configuration, the MRS for these fields are likely to be somewhere between $5\arcsec$ to $12\arcsec$. In Fig.~\ref{fig2}, we position an MRS of $5\arcsec$, $8\arcsec$, and $12\arcsec$ on the $\Delta$-variance spectra in order to assess in how many fields is $L_{\rm hub}$ larger than any of these assumed MRS values. A visual inspection of Fig.~\ref{fig2} shows that adopting a conservative value of $5\arcsec$ for the MRS for all fields implies that 4-5 regions have their corresponding values of $L_{\rm hub}$ larger than this value of the MRS. These are G353.41, W43-MM3, G012.080, G333.60, and marginally G337.92. For the other regions, the values of $L_{\rm hub}$ are well below $5\arcsec$, up to a factor of $\approx 5$ smaller. For an MRS of $8\arcsec$, the number of fields for which this is the case is reduced to three (W43-MM3, G012.80, and G333.60), and for an MRS of $12\arcsec$, only two regions have their values of $L_{\rm hub}$ coincide with the value of the MRS (G01280 and G333.60).

While interferometric observations cause a partial loss of the signal on scales larger than the maximum recoverable scale (MRS), and the effects of this dimming increase with spatial scales that are increasingly larger, this loss of emission on the largest scales does not imply that such large structures cannot be detected in the $\Delta$-variance spectrum. For dimmed structures to be detected, some conditions have to be fulfilled. The first one is that the large structure should be coherent enough to be recognized as a single structure, even if it contains dense substructures (i.e., cores) within it. The second condition is that such large structures should be sufficiently massive and dense in order to have a relevant contribution for star formation within the cloud. The potential non-detection of large and coherent structures that are already too dim before being filtered out by the interferometer do not impact the correlation between the size of over-dense star forming regions and the star formation rate (SFR) indicators. On the other hand, such large structures that could be filtered out by the interferometer can not be as dense as the star forming clumps such as those seen in the ALMA-IMF field. The reasoning behind this is that such large clumps (larger than those currently observed) would host a large number of dense cores with a well structured geometry and they should be easily detected. In some of the most evolved regions, this would also imply an increased signature of stellar feedback. In the ALMA-IMF fields, both dense cores and feedback effects are solely observed inside or in close proximity of the presently detected massive clumps. This leaves the possibility of moderately dense clumps that, once dimmed, could potentially appear smaller as only their denser inner regions are detected above the noise level. In Appendix A, we explore this possibility using simple idealized tests and show that the entire size of moderately over-dense clumps can still be detected in the $\Delta$-variance spectrum even when the clumps are dimmed by a factor of $75\%$. This is not a hard limit and the real value is likely to depend on both the local and global background with the respect to the peak densities of the clumps, their size, and density profiles. For the ALMA-IMF sample, combining images obtained with the 12m array with others obtained with the 7m array can help recover the entire signal up to larger scales. However, and while observations do exist with the 7m array for the ALMA-IMF regions, combining them with the 12m array observations has proved to be problematic for most regions (Ginsburg et al. 2022). An example where the image combination is successful (G328.25, B3 band) is displayed in Figure H.10 of their paper. In this case, more substructure can be observed in the combined image than in the 12m image. Yet, it is evident that the extended large scale substructure is of low intensity and would not result in any peak the $\Delta$-variance spectrum. Furthermore, the size of the over-dense clump in G328.25 is also virtually unchanged in the combined image. In Appendix B, we confirm this conclusion by comparing the $\Delta$-variance spectrum of both maps.

Figure~\ref{fig4} displays the value of $L_{\rm hub}$ plotted against the $\left(S_{\rm 1.3 mm}^{\rm cloud}/S_{\rm 3mm}^{\rm cloud}\right)$ ratio (left panel) and $\Sigma_{\rm H\alpha{41}}^{\rm free-free}$ (right panel). A strong anti-correlation is clearly observed between the $\left(S_{\rm 1.3 mm}^{\rm cloud}/S_{\rm 3mm}^{\rm cloud}\right)$ ratio and $L_{\rm hub}$ (Spearman $\rho$ coefficient of $-0.62$ with a significance for the deviation from zero of $0.012$) and a strong correlation is observed between the latter and $\Sigma_{\rm H\alpha{41}}^{\rm free-free}$ (Spearman $\rho$ coefficient of $0.53$ and a significance of deviation from zero of $0.03$). In calculating the correlation coefficient, we have again excluded the G328.25 region as its measured value of $L_{\rm hub}$ is very close to the resolution limit.  

If the structures within each clouds were the consequence of compression by turbulent motions that are solely cascading from larger scales, one would expect the values of $\alpha$ to be nearly identical among the ALMA-IMF clouds and be independent of the evolutionary stage. This would be particularly the case for the less evolved clouds which are not affected by stellar feedback emanating from within the dense clumps or from their direct vicinity. Compressive motions due to feedback can alter the properties of turbulence in some clouds and alter the clouds structure on small scales (Federrath et al. 2008; Luisi et al. 2021; Bonne et al. 2022). Gravity, particularly around the central regions of the clouds, is likely to be a major player in both generating infall-induced turbulence and in shaping the clouds' structure (Dib et al. 2007; Ballesteros-Paredes et al. 2011; V\'{a}zquez-Semadeni et al. 2017; Xu \& Lazarian 2020). Thus, the large scatter in the values of $\alpha$ is likely an indication that the self-similar structures that are observed in the $1.3$ mm emission maps of most regions are not the result of turbulence cascading from larger scales alone and that it may contain a significant contribution from infall motions due to the self-gravity of the clouds as well as from feedback processes in some of the evolved clouds. The existence of a significant correlation between $L_{\rm hub}$ and $\left(S_{\rm 1.3}^{\rm cloud}/S_{\rm 3 mm}^{\rm cloud}\right)$ and of a significant anti-correlation between $L_{\rm hub}$ and $\Sigma_{\rm H\alpha{41}}^{\rm free-free}$ also points to a significant role for gravity in driving the growth of the protocluster clump with continued gas accretion.

\section{CONCLUSIONS}\label{conc}

We calculated the $\Delta$-variance spectra of all 15 protocluster forming clouds in the ALMA-IMF survey. The shapes of the spectra resemble those found earlier for closer massive star forming regions and all of them display a perturbed self-similar regime on small scales followed by a prominent bump at larger scales. This feature in the spectra is indicative of the presence of over-dense structures which are the sites of ongoing and future massive star and cluster formation. The sizes of those regions, $L_{\rm hub}$, is correlated with the surface density of the $\rm H41\alpha$ free-free emission and anti-correlated with the ratio of the $1.3$ mm to $3$ mm continuum emission fluxes $\left(S_{\rm 1.3 mm}^{\rm cloud}/S_{\rm 3 mm}^{\rm cloud}\right)$. Since smaller values of $\left(S_{\rm 1.3 mm}^{\rm cloud}/S_{\rm 3 mm}^{\rm cloud}\right)$ and larger values of $\Sigma_{\rm H41\alpha}^{\rm free-free}$ correspond to more advanced evolutionary stages of the protocluster forming clumps, this indicates that the size of the densest regions in protocluster forming clouds is a direct indicator of its evolutionary stage, with larger clumps being associated with a more advanced evolutionary stage. We do not find any correlation between the value of the exponent of the power law that characterize the self-similar regime with either $\Sigma_{\rm H41\alpha}^{\rm free-free}$ or $\left(S_{\rm 1.3 mm}^{\rm cloud}/S_{\rm 3 mm}^{\rm cloud}\right)$. These results indicate that turbulent motions which would set a self-similar regime in the clouds are not only those inherited by a cascade from larger scales, but have a significant contribution from gravity-driven turbulence, with a potential contribution from feedback effects in the most evolved clouds. The correlation between $L_{\rm hub}$ and evolutionary indicators (here the $\left(S_{\rm 1.3 mm}^{\rm cloud}/S_{\rm 3 mm}^{\rm cloud}\right)$ ratio and $\Sigma_{\rm H41\alpha}^{\rm free-free}$) is a clear indication that the densest regions of the clouds are dominated by gravity which must be playing a major role in regulating their size and mass growth. 
                                                                             
\section{Acknowledgments}

I thank the referees for useful feedback that helped improve his paper. I also thank Adam Ginsburg for very useful clarifications on aspects of the ALMA-IMF data and Volker Ossenkopf-Okada for helpful insight into the calculation of the $\Delta$-variance spectra. 

\section{Data availability}
The data underlying this article will be available upon reasonable request.

{}

\begin{appendix}

\section{The detection of dimmed structures with the $\Delta$-variance spectrum}\label{appa}

\begin{figure*}
\centering
\includegraphics[width=0.42\textwidth] {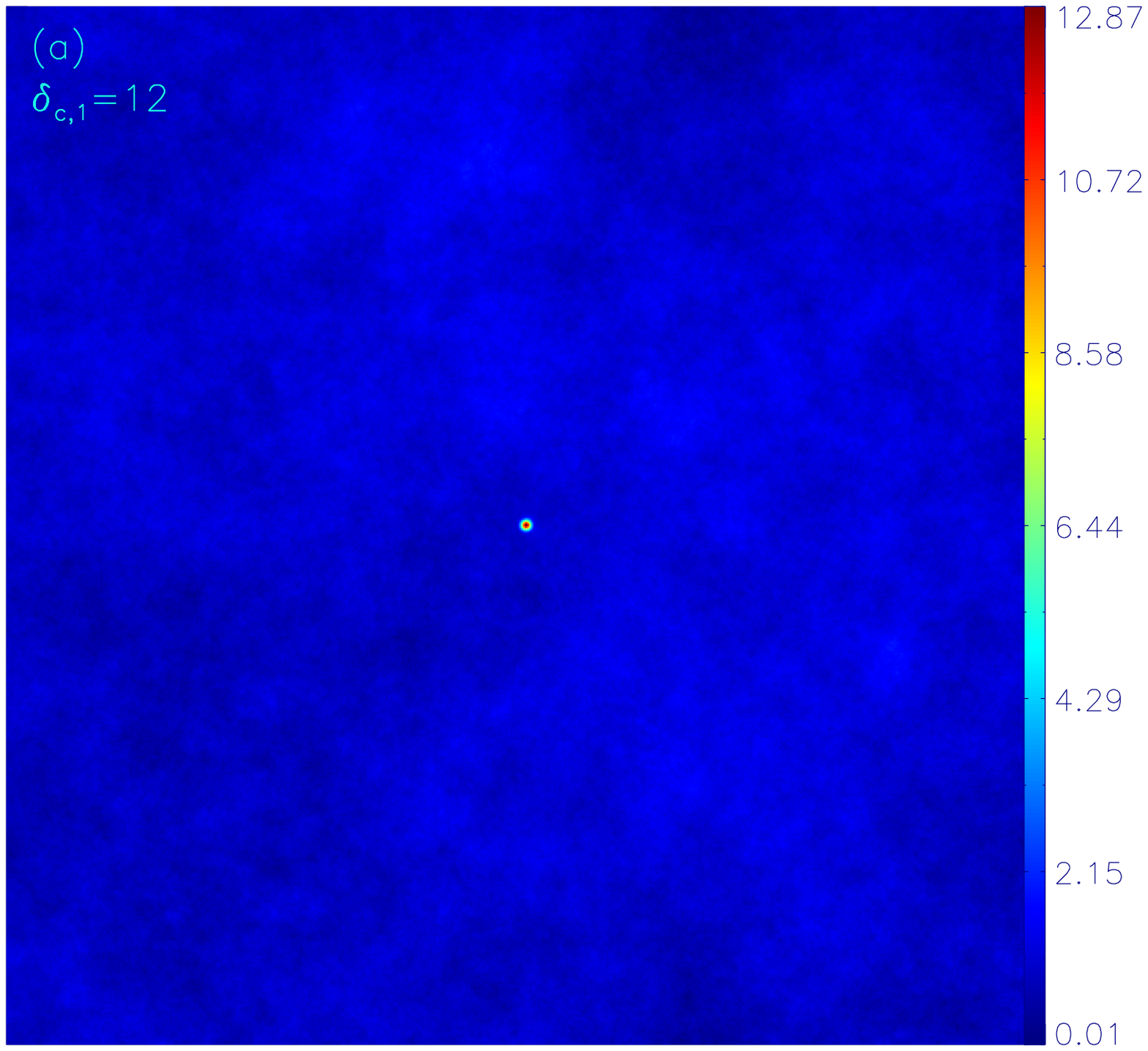}
\hspace{0.8cm}
\includegraphics[width=0.42\textwidth] {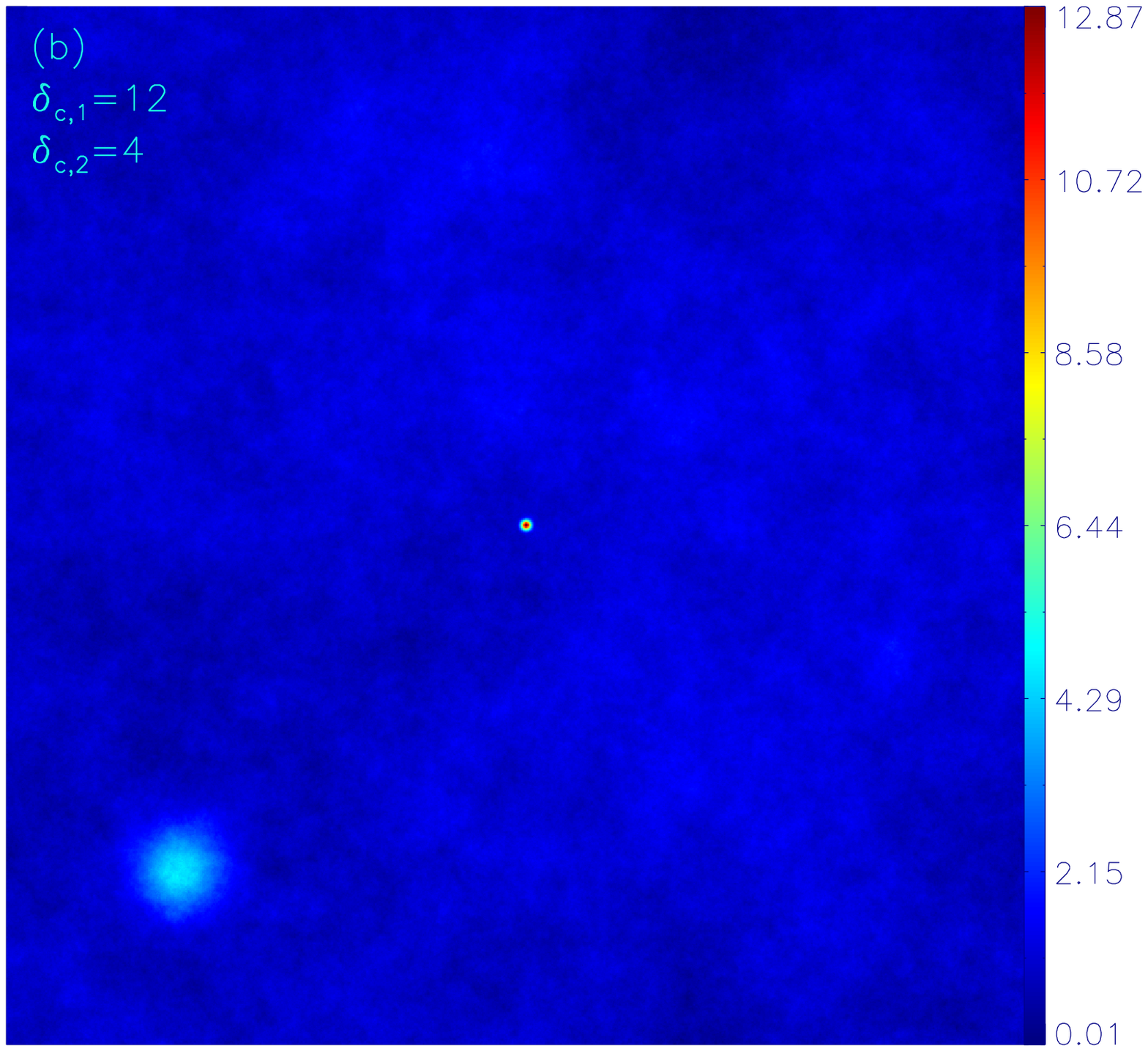}\\
\vspace{0.3cm}
\includegraphics[width=0.42\textwidth] {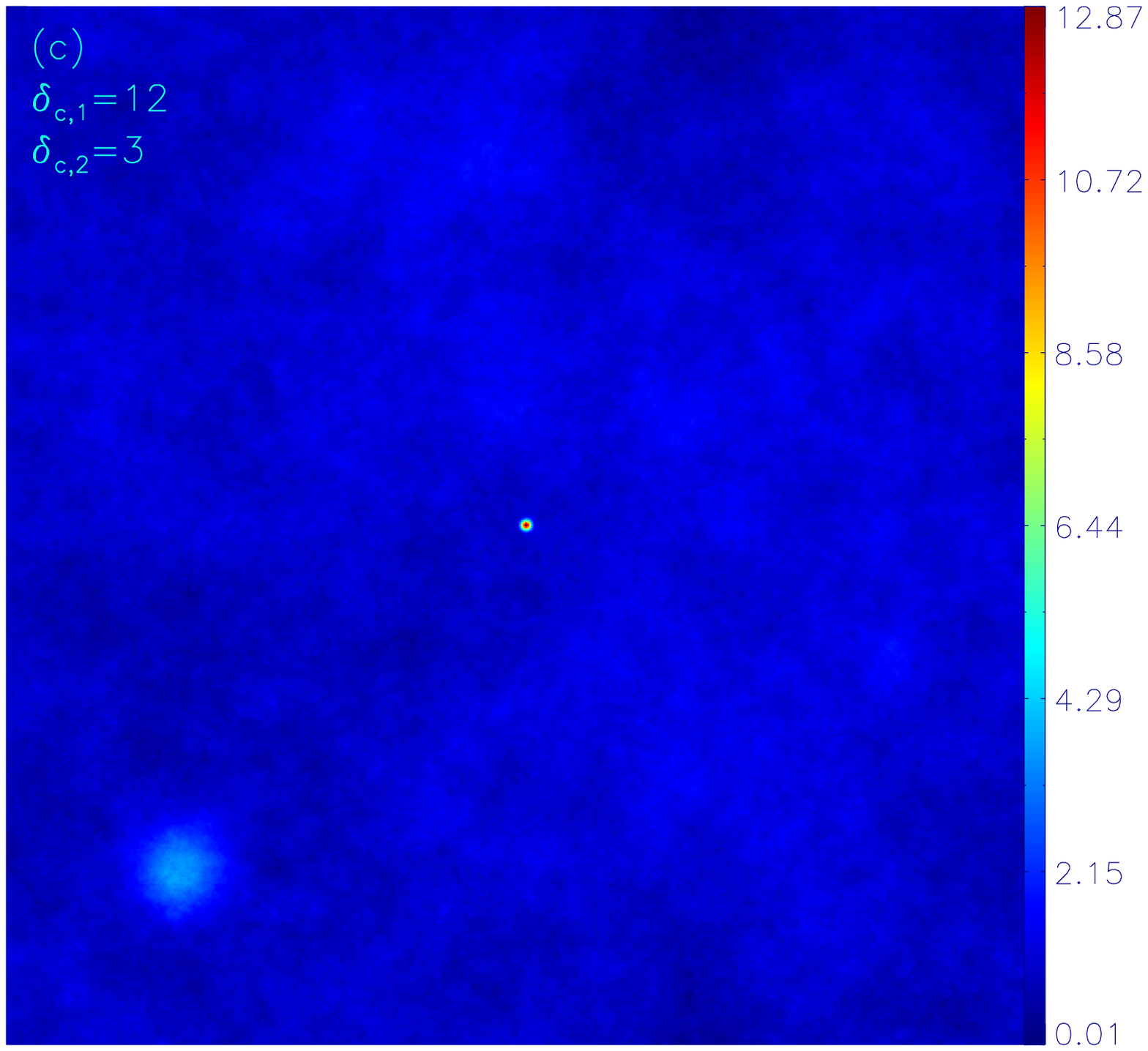}
\hspace{0.8cm}
\includegraphics[width=0.42\textwidth] {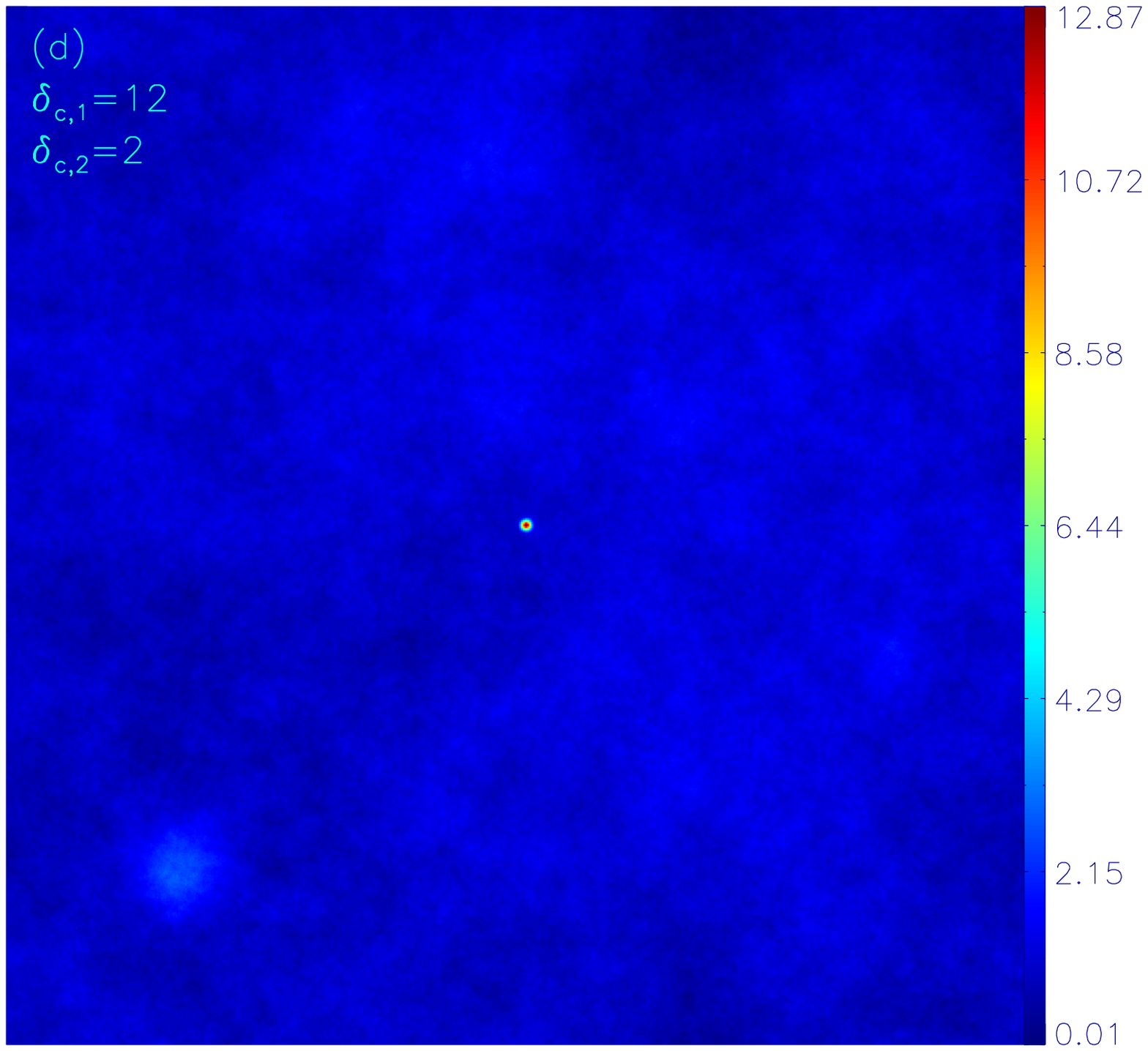}\\
\vspace{0.3cm}
\includegraphics[width=0.42\textwidth] {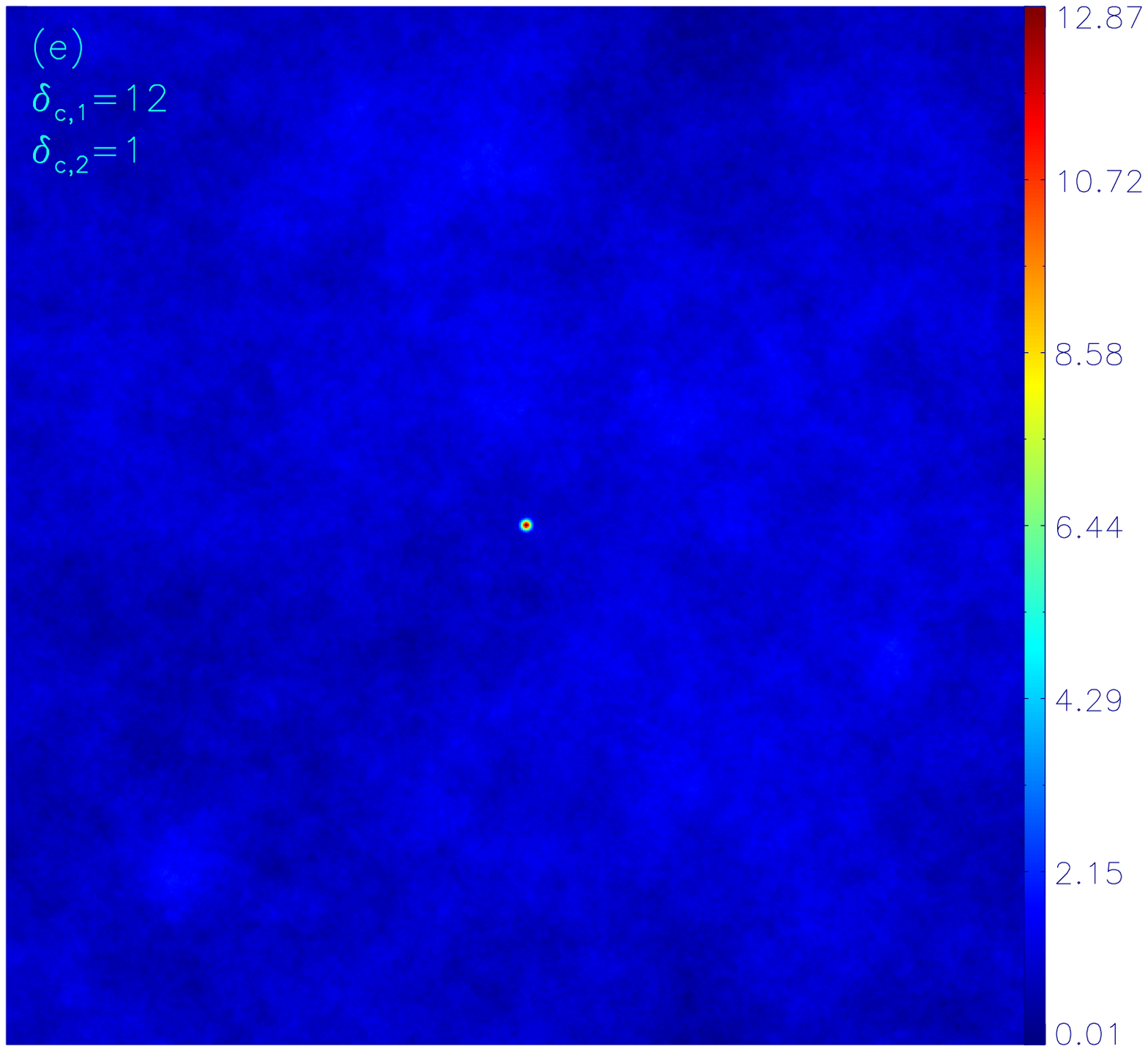}
\hspace{0.8cm} 
\includegraphics[width=0.42\textwidth] {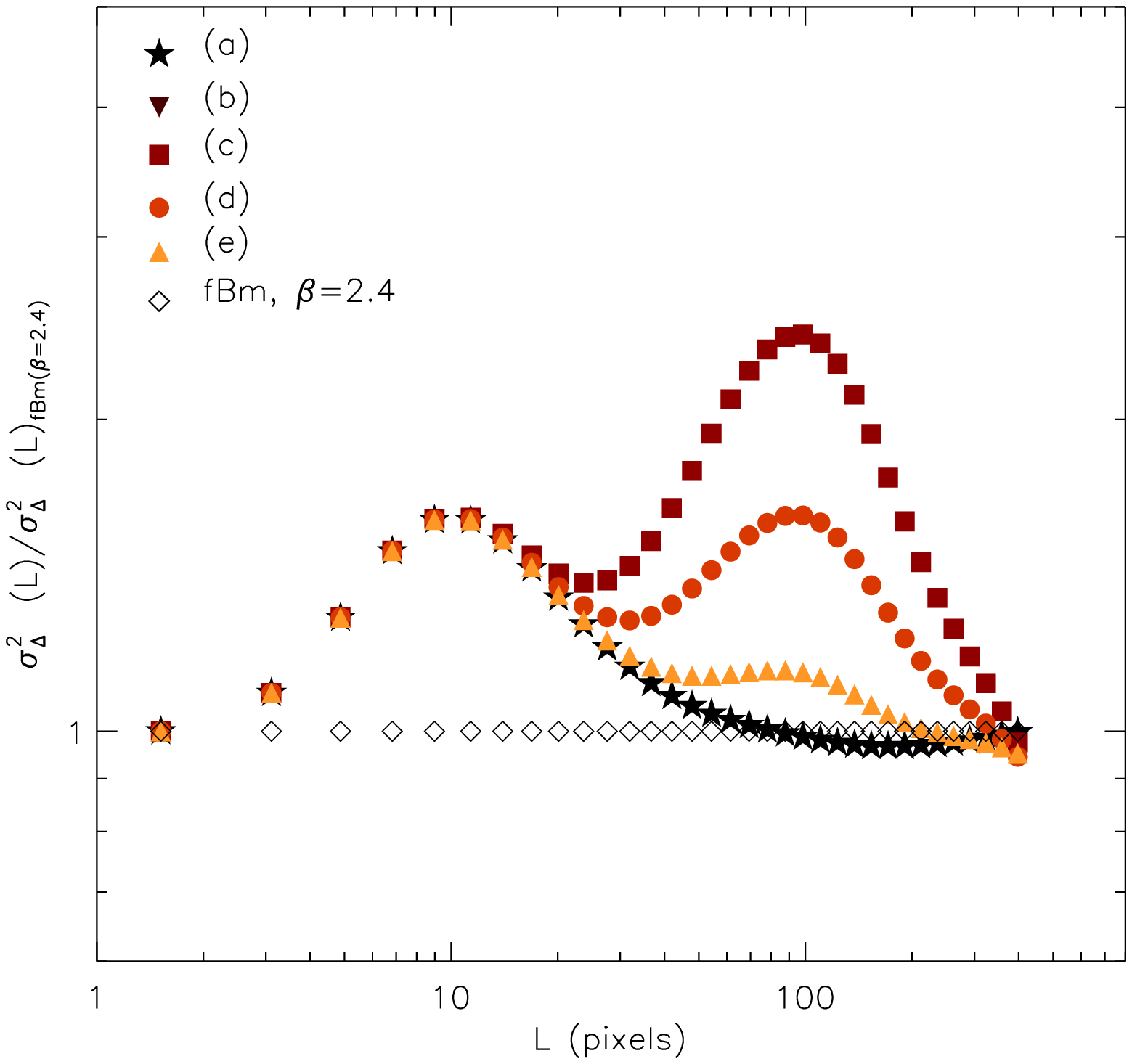}
\caption{2D Gaussian structures injected on top of an fBm image with $\beta=2.4$. The 2D Gaussian functions have an aspect ratio of unity. In panel (a), a clump with a standard deviation of $3.3$ pixels in both direction is injected on top of the fBm image. The clump has an overdensity between its peek and the average value of the fBm image of $\delta_{c,1}=12$.  In panel (b), a larger clump (standard deviation of 30 pixels in both direction) is added and has a value of $\delta_{c,2}=4$. In panels (c), (d), and (e), the larger clump is dimmed by factors of $25\%$, $50\%$, and $75\%$ from its original value. The bottom right panel displays the corresponding $\Delta$-variance functions calculated for each case and they are normalized by that of the fBm.}
\label{figapp1}
\end{figure*}

Here, we present a suite of simple and idealized tests with the aim of investigating to which extent large clumps that would be dimmed in interferometric observations remain detectable in the $\Delta$-variance spectrum. As a background, we assume a $1000\times1000$ pixels fBm with an exponent of $\beta=2.4$ similar to the one used in Dib et al. (2020). The $\Delta$-variance spectrum of this fBm is a power law with an exponent of $\alpha=0.4$. In the first instance, we consider a single over-dense clump overlayed on top of the same fBm (panel (a) in Fig.~\ref{figapp1}). The clump is modeled as a 2D Gaussian with a density contrast between the peak and the mean value of the background of $\delta_{c,1}=12$. The standard deviation of the Gaussian in each direction is $\sigma_{1}=3.3$ pixels. The map is scale free, and this clump could be thought of as being representative of some of the dense clumps that are presently detected in the ALMA-IMF fields with a size of a few arcsec. The $\Delta$-variance spectrum of this map is calculated and displayed in the lower right panel of Fig.~\ref{figapp1}. The spectrum is divided by that of the fBm in order to clearly identify the signature of the clump. The $\Delta$-variance clearly displays a peak at a scale of $\approx 10$ pixels which is $\approx 3 \sigma_{1}$. We remind that the amplitude of the bump scales as $A \delta_{c}^{2}$ where $A$ is the surface area of the clump or collective surface if several clumps are present (see Dib et al. 2020). As such, a clump that is two times brighter will create a more prominent bump for a given size than a clump that is two times larger at the same level of brightness.

We would like to test the situation when a second, larger clump or structure exists in the map alongside the first one. Translating in terms of the ALMA-IMF images, and as explained in \S.~\ref{res}, the second clump cannot be denser or as dense as the first clump that are presently detected. This is because such a bright large clump would have been selected as the primary target of the observations (and not the smaller clump), and secondly and most importantly, the existence of such a large bright clump implies the observations of many dense cores that are spatially correlated on a spatial area that is much larger than what is presently observed since dense cores cannot be filtered out by the interferometric observations. In contrast, large dim clouds with a contrast that is a few percent or a few tens of percent above the global background (i.e., $\delta_{c,2} < 2$) are not extremely interesting, even though they are likely to be present in all maps (see \S.~\ref{appb}). Albeit these dim large clumps are likely to be filtered out in the 12m configuration observations, their detection, or lack of it, is unlikely to affect the position of the peak since they do not create substantial variance on the map on their physical scale. 

The remaining possibility is the existence of large structures of intermediate brightness ($\delta_{c} >2$ but smaller than $\delta_{c,1}$). By varying the surface area of the second clump and its $\delta_{c}$, we consider a second clump that will have a bump in the $\Delta$-variance spectra that is lower than the first clump. However, this case, although likely close to reality (see \S.~\ref{appb} and the $\Delta$-variance of the Cygnus-North region in Dib et al. (2020) is not informative since we are assuming already that the peak corresponds to the size of the first clump. Instead, we consider a "problematic" case in which the amplitude of the second clump is larger than that of the first one by carefully selecting its surface and value of $\delta_{c,2}$ and ask the question whether this second clump can still be detected as a second bump in the $\Delta$-variance spectrum even when its amplitude is reduced and falls below that of the first clump. We consider a second clump with a density contrast of $\delta_{c,2}=4$ between its peak value and the mean value of the background. The second clump is also modeled as a 2D Gaussian with a standard deviation of $\sigma_{2}=30$ pixels in both directions (panel (b) in Fig.~\ref{figapp1}). The $\Delta$-variance spectrum corresponding to this situation is displayed in the lower right panel of Fig.~\ref{figapp1}. The $\Delta$-variance spectrum clearly displays two separate peaks that correspond to the smaller clump as well as the less dense, larger clump. The peaks correspond to the full size of the clumps. We now attenuate the emission of the larger clump by $25\%$, $50\%$, and $75\%$ from its original value (panels (c), (d) and (e) respectively). For simplicity, large scale structures found in the underlying fBm are not dimmed as these are already of lower intensity than both clumps. The corresponding $\Delta$-variance for these cases are again displayed in the lower right panel of Fig.~\ref{figapp1}. With an increased dimming of the large clump, the amplitude of the corresponding bump it generates in the $\Delta$-variance spectrum is reduced. However, it can still be measured accurately even when the amplitude of the 2D Gaussian is reduced by $75\%$ from its original value. In this case, the $75\%$ threshold corresponds to a peak emission of the 2D Gaussian of the large clump that is equal to one time the average density of the background. Below this value, the second peak corresponding to the large clump is blurred with the background. This is however not a hard limit, and the threshold could be either lower or higher, depending on a number factors such as the size of the clump, its density profile, and the emission levels of the local background. The idealized test presented here shows that while interferometric observations may decrease the amplitude of clumps or structures of intermediate brightness, their detection and the measurement of their size is still possible, even when being partially filtered out. In the $\Delta$-variance spectra of the different ALMA-IMF regions, it is not obvious at all that we can see any secondary peak apart from that associated with the main clump. This implies that the largest structures that exist in the maps are already dim ($\delta_{c} < 2$ or less) as is exemplified by the case of the G328.25 region for which a merged image from the 12m and 7m configuration exists and in which flux is recovered for spatial scales that are larger than those of the 12m map alone, and as is observed in the {\it Herschel} satellite images in which no flux is lost at large scales (e.g., Kumar et al. 2020).

\section{Comparing the $\Delta$-variance spectrum of the 12m image to the combined 12m+7m merged image}\label{appb}

\begin{figure}
\begin{center}
\includegraphics[width=0.9\columnwidth] {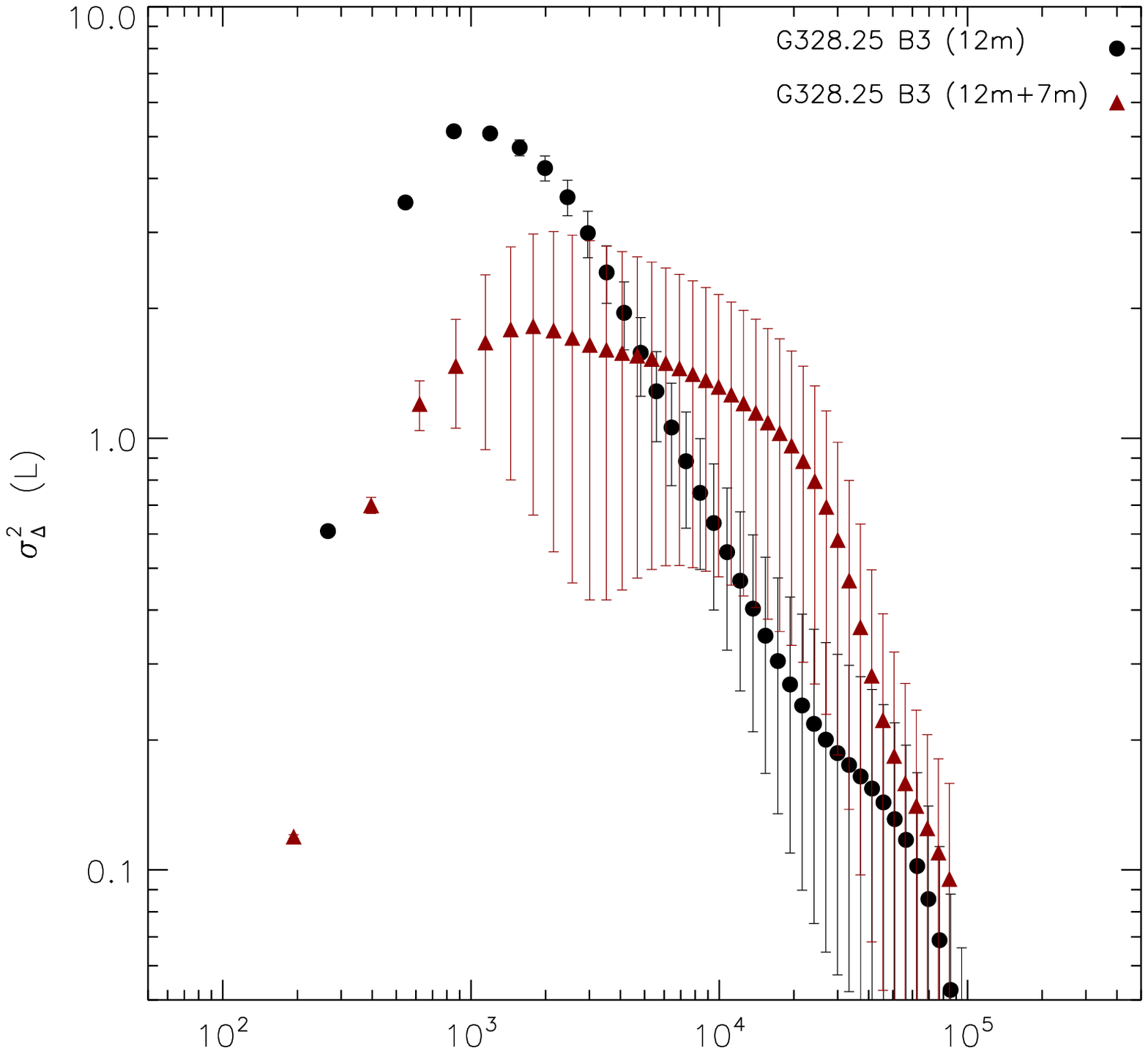}
\vspace{0.75cm}
\caption{$\Delta$-variance spectra of the G328.25 3mm image for the 12m only configuration and the 3mm 12m+7m combined image. Both images are those displayed in Figure H.10 in Ginsburg et al. (2022).}
\label{figapp2}
\end{center}
\end{figure}  

Ginsburg et al. (2022) discussed in detail the difficulties pertaining to the construction of merged images of the different regions from the 12m and 7m configurations. They also discuss how the combination of the 12m and 7m images increases the noise levels. They succeeded in deriving such a map only for one region and one band, namely the 3mm band of G328.25 (Figure H.10 in their paper). We have calculated the $\Delta$-variance spectrum of both maps (the tt0.pcor version) and the results are displayed in Fig.~\ref{figapp2}. The comparison of the two spectra confirms what is observed in the combined image, namely that there are some larger structures present in the map. These structures are of intermediate size (i.e., a few times larger than the central clump), and of low mean emission. This translates in the $\Delta$-variance spectrum into more variance at around these scales, resulting in a wider spectrum than that corresponding to the 12m case. However, what can also be observed is that the position of the peak is roughly the same in both cases and it only shifts by about $50\%$ to larger values, namely from around $\approx 1000$ au to $\approx 1500$ au. What is also important to note is that since the noise levels are higher in the combined image, this leads to a $\Delta$-variance spectrum with larger error bars. In the absence of combined images for the other regions, it is difficult to ascertain whether variations in the positions of the peak in the $\Delta$-variance spectrum will be of the same order in other regions. Should that be the case, the correlation that is found between $L_{\rm hub}$ and the SFR indicators (in this work the $\left(S_{\rm 1.3 mm}^{\rm cloud}/S_{\rm 3mm}^{\rm cloud}\right)$ ratio and $\Sigma_{\rm H\alpha{41}}^{\rm free-free}$) should not be affected in a radical way. If the $L_{\rm hub}$ estimates for all regions were to increase by a few tens of percent, this will change the shape of the correlation between $L_{\rm hub}$ and the SFR indicators, but without suppressing it since, even when excluding the G328.25 region, the values of $L_{\rm hub}$ for the other regions vary by almost one order to magnitude.

\end{appendix}

\label{lastpage}

\end{document}